\documentclass[aps,twocolumn,nofootinbib,longbibliography]{revtex4-1}

\usepackage{enumerate}
\usepackage{fancyhdr}
\usepackage{amsmath,amsfonts,amssymb}
\usepackage[colorlinks,citecolor=blue,linkcolor=blue,urlcolor=blue]{hyperref}

\usepackage{color}
\definecolor{red}{rgb}{1,0,0}
\definecolor{gre}{rgb}{0,0.6,0}
\definecolor{blu}{rgb}{0,0,1}

\usepackage[pdftex]{graphicx}
\usepackage{mathrsfs}
\def\be{\begin{equation}}
\def\ee{\end{equation}}

\def\r{R}

\begin{document}

\title{Black hole fireworks: quantum-gravity effects outside the horizon spark black to white hole tunneling}

\date{Fourth of July, 2014}

\author{Hal M. Haggard}
\email{hal.haggard@cpt.univ-mrs.fr}
\author{Carlo Rovelli}
\email{rovelli@cpt.univ-mrs.fr}
\affiliation{
Aix-Marseille Universit\'e and Universit\'e de Toulon, 
CPT-CNRS, Luminy, F-13288 Marseille
}%

\begin{abstract}
\noindent We show that there is a classical metric satisfying the Einstein equations outside a finite spacetime region where matter collapses into a black hole and then emerges from a white hole. We compute this metric explicitly.  We show how quantum theory determines the (long) time for the process to happen.  A black hole can thus quantum-tunnel into a white hole. For this to happen, quantum gravity should affect the metric also in a small region outside the horizon: we show that contrary to what is commonly assumed, this is not forbidden by causality or by the semiclassical approximation, because quantum effects can pile up over a long time. This scenario alters radically the discussion on the black hole information puzzle. 
\end{abstract}

\maketitle

%\tableofcontents

\section{What happens at the center of a black hole?}

Black holes have become conventional astrophysical objects. Yet, it is surprising how little  we know about what happens inside them.  Astrophysical observations indicate that general relativity (GR) describes well the region surrounding the horizon (see e.g. \cite{Narayan2013}); it is plausible that also a substantial region inside the horizon is well described by GR.  But certainly classical GR fails to describe Nature at small radii, because nothing prevents quantum mechanics from affecting the high curvature zone, and because classical GR becomes ill-defined at $r\!=\!0$ anyway. The current tentative quantum gravity theories, such as loops and strings, are not sufficiently understood to convincingly predict what happens in the small radius region, so we are quite in the dark: what ultimately happens to gravitationally collapsing matter? Does it emerge into a baby universe (as in Smolin's cosmological natural selection \cite{Smolin1997})? Does it vanish mysteriously ``into a deep interior where space and time and matter as we know them lose their meaning"?\,... 

There is a less dramatic possibility, which we explore in this paper:  when matter reaches Planckian density, quantum gravity generates sufficient pressure to counterbalance the matter's weight, the collapse ends, and matter bounces out.  A collapsing star might avoid sinking into $r\!=\!0$ much as a quantum electron  in a Coulomb potential does not sink all the way into $r\!=\!0$.  The possibility of such a Planck star phenomenology has been considered by numerous authors \cite{Frolov:1979tu,Frolov:1981,Stephens1994,Mazur:2004,Ashtekar:2005cj,Hayward2006,Hossenfelder:2010,frolov:BHclosed,Bardeen2014,Rovelli2014,Barrau2014}. The picture is similar to Giddings's remnant scenario \cite{Giddings:1992}, here with a macroscopic remnant developing into a white hole. Here we study if it is compatible with a realistic effective metric \emph{satisfying the Einstein equations everywhere outside the quantum region}.  

Surprisingly, we find that such a metric exists: it is an exact solution of the Einstein equations everywhere, including inside the Schwarzschild radius, except for a \emph{finite}---small, as we shall see---region, surrounding the points where the classical Einstein equations are likely to fail. It describes in-falling and then out-coming matter. 

A number of indications make this scenario plausible. H\'aj'\v{c}ek and Kiefer \cite{HAJICEK2001} have studied the dynamics of a null spherical shell coupled to gravity.  The classical theory has two disconnected sets of solutions: those with the shell in-falling into a black hole, and those with the shell emerging from a white hole. The system is described by two variables and can be quantized exactly.  Remarkably, the quantum theory connects the two sectors: a wave packet representing an in-falling shell tunnels (undergoing a quantum ``bounce") into an expanding wave packet. H\'aj'\v{c}ek and Kiefer do not write the effective metric that describes this process; here we do.  

A similar indication for the plausibility of this scenario comes from loop cosmology: the wave packet representing a collapsing universe tunnels into a wave packet representing an expanding universe  \cite{Ashtekar2006}. Again, the quantum theory predicts tunnelling between two classically disconnected sets of solutions: collapsing and expanding.  In this case, an effective metric is known that describes the full process \cite{Ashtekar:2006es}, and indeed does so in a surprisingly accurate way \cite{Rovelli2013e}; it satisfies the classical Einstein equations everywhere except for a small region where quantum effects dominate and the classical theory would become singular. 

The technical result of the present paper is that such a metric exists for a bouncing black to white hole. It solves the Einstein equations outside a finite radius and beyond a finite time interval.  Its existence shows that it is possible to have a black hole bouncing into a white hole \emph{without affecting spacetime at large radii}. The quantum region extends just a bit outside $r=2m$ and has short duration. The metric describes also the region inside $r=2m$. A distant observer sees a dimming, frozen star that reemerges, bouncing out after a very long time (computed below), determined by the star's mass and Planck's constant. 

Two natural obstacles have made finding this metric harder.  The first is its technical complication: the metric we find is locally isometric to the Kruskal solution (outside the quantum region), but it is \emph{not} a portion of the Kruskal solution. Rather, it is a portion of a double cover of the Kruskal solution, in the sense that there are \emph{distinct} regions isomorphic to \emph{the same} Kruskal region.  This is explained in detail below, and is the technical core of the paper. 

But the larger obstacle has probably been a widespread uncritical assumption: that Nature should be well approximated by \emph{one and the same} solution of the classical equations in the \emph{entire} region where curvature is small.  This is a prejudice because it neglects the fact that small effects can pile up in the long term. If a perturbation is small, then the true dynamics is well approximated by an unperturbed solution \emph{locally}, but not necessarily globally:  a particle subject to a weak force $\epsilon\, F$ where $\epsilon\!\ll\! 1$ moves as $x=x_o+v_0t + \frac12 \epsilon\, F t^2$.  For any small time interval this is well approximated by a motion at constant speed, namely a solution of the unperturbed equation; but for any $\epsilon$ there is a $t\sim 1/\sqrt{\epsilon}$ long enough for the discrepancy between the unperturbed solution and the true solution to be arbitrarily large.   

Quantum effects  can similarly pile up in the long term, and tunneling is a prime example: with a very good approximation, quantum effects on the stability of a single atom of Uranium  238 in our lab are completely negligible. Still, after 4.5 billion years the atom is likely to have decayed. Outside a macroscopic black hole the curvature is small and quantum effects are negligible, today. But over a long enough time they may drive the classical solution away from the exact global solution of the classical GR equations.  After a sufficiently long time, the hole may tunnel from black to white.   This is the key conceptual point of this paper, and is discussed in detail in Section II, where we show, in particular, that there is no causality violation involved in this effect. 

Importantly, the process is very long seen from the outside, but is very short for a local observer at a small radius.  Thus, classical GR is compatible with the possibility that a black hole is a (quantum) bouncing star seen in extreme slow motion.  The bounce could lead to observable phenomena \cite{Rovelli2014} whose phenomenology has been investigated in \cite{Barrau2014}.

Anticipating what we find below, quantum effects can first appear at a radius 
\be
   r\sim \frac76\ 2m
\ee
after an (asymptotic) proper time of the order
\be
   \tau\sim \frac{m^2}{l_{P}},
\ee
where $l_{P}$ is the Planck length (we use units where the speed of light and Newton's constant are $c=G=1$). This time is very long for a macroscopic black hole (it is equal to the Schwarzschild time multiplied by the ratio between the mass of the collapsing object and the Planck mass; this is huge for a star), but is shorter than the Hawking evaporation time, which is of order $m^3$. Therefore the possibility of the bounce studied here affects radically the discussion about the black hole information puzzle. 

A word about the relation between our results and the firewall discussion \cite{Almheiri:2012rt} is thus perhaps useful.  The firewall argument indicates that under a certain number of assumptions ``something strange" appears to have to happen at the horizon of a macroscopic black hole.  Here we point out that indeed it does,  \emph{independently from the Hawking process}, but it is a less dramatic phenomenon than expected: the spacetime quantum tunnels out of the black hole and this can happen without violating causality because over a long stretch of time quantum gravitational effects can accumulate outside the horizon and modify the metric beyond the apparent horizon. 

In the next section, Section \ref{ball}, we give a preliminary discussion of the quantities in play by studying a simple situation where no actual horizon develops, but a parameter can be tuned to approach a situation with horizon.  This allows us to discuss the timing and the location of the appearance of quantum effects outside the horizon. In Section \ref{III} we define precisely the problem we want to solve, namely the characteristic of the metric we are seeking.  This metric is constructed in Section \ref{metric}. In Section \ref{time} we show that the process has short duration seen from the inside and long duration seen from the outside.  In Section \ref{constants} we determine all the constants left free in the definition of the metric.  Section \ref{conclusion} summarises our results and discusses how it could be connected to a full fledged theory of quantum gravity. 

\section{Preliminary discussion: the crystal ball}\label{ball}

Consider a ball of radius $a$ with perfectly reflective surface, a mass negligible for the present discussion, at rest in flat space.  Consider an incoming shell of light, with total energy $m$ centered on the center of the crystal ball, coming in from infinity.   What happens next?  

\begin{figure}[t]
\includegraphics[height=7cm]{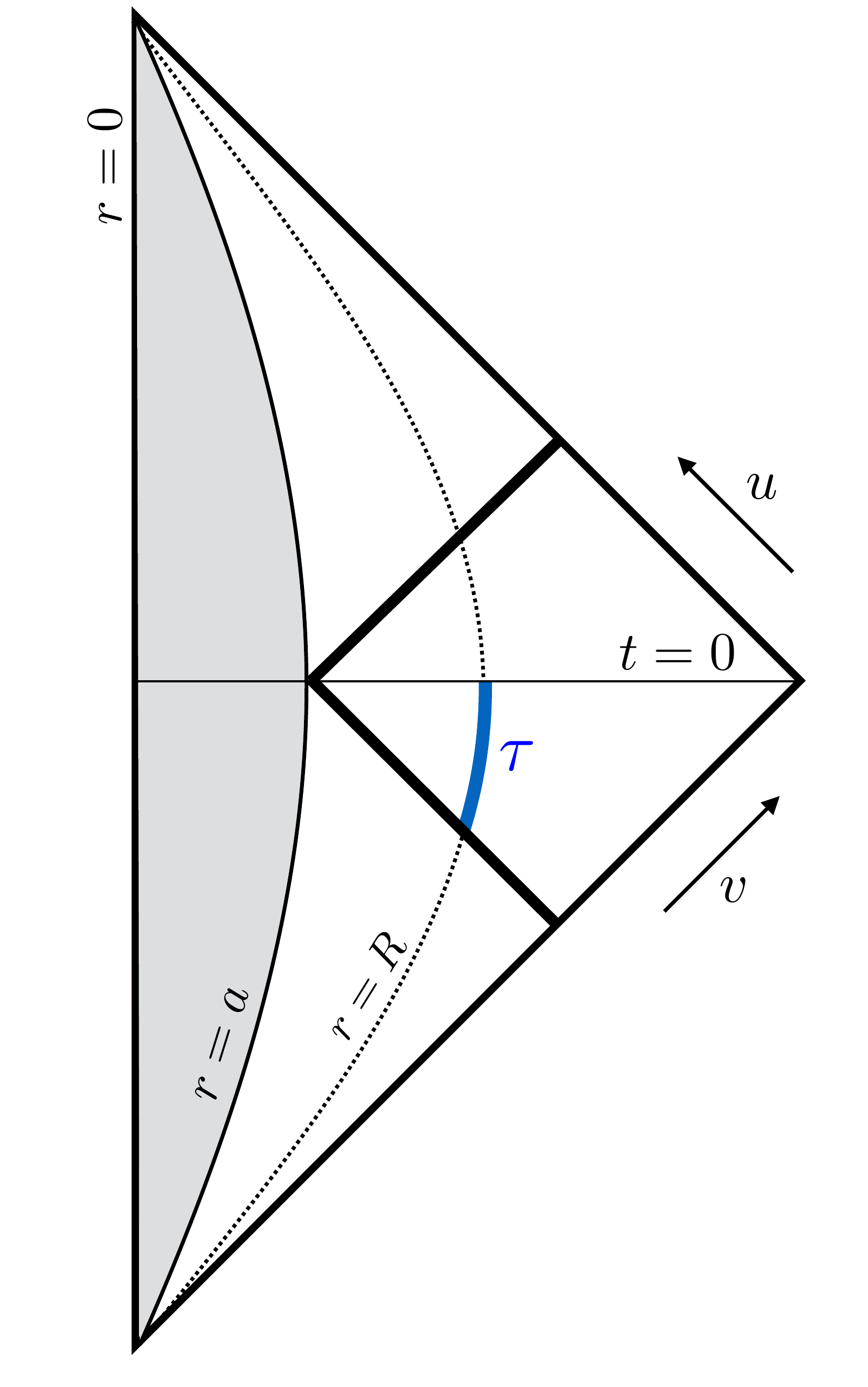}
\vspace{-1em}
\caption{The crystal ball is in grey, the thick lines represent the bouncing shell of light. The dotted line is the observer and the bounce time $\tau$ is indicated in blue.}
\label{uno}
\end{figure}

The answer depends on the relation between $m$ and $a$.  Suppose first that $a\!\gg\! 2m$. Then we are in a non-relativistic regime. Outside the collapsing shell the metric is just the (large-radius part of the) Schwarzschild metric of mass $m$.  The shell moves in until $r\!=\!a$ then bounces out.  In a two-dimensional conformal diagram, the situation is illustrated in Figure \ref{uno}.

Consider an observer sitting at a reference radius $R$. He will measure a proper time $2\tau_R$ between the moment the shell passes him incoming and the moment it passes him outgoing. We call $\tau_R$ the ``bounce time", seen by the observer at $R$. Let us study how it depends on $a$ and $R$.  As long as $a\!\gg\!2m$, we can neglect relativistic effects, and we have trivially $\tau_R=R-a$: the time it takes light to reach the mirror.  If we decrease the radius $a$ of the ball, the time $\tau$ increases.  When $a$ becomes of the order of $2m$ (but still $a\!>\!2m$), we enter a general relativistic regime, and we must take this into account; the dependence of $\tau$ on $a$ and $R$ becomes more interesting. The metric outside the shell is Schwarzschild (the region to the right in the conformal diagram). In null Kruskal coordinates this is 
\be
ds^2=-\frac{32m^3}{r}e^{-\frac{r}{2m}} du dv + r^2 d\Omega^2,
\ee
where $d\Omega^2$ is the metric of the unit sphere and $r(u,v)$ is determined by 
\be
 \left(1-\frac{r}{2m}\right)e^{\frac{r}{2m}}=uv.
 \label{r}
\ee
If we place the bounce at $(u\!+\!v) \!=\!0$, which corresponds with $t=0$, the trajectory of the incoming shell, an incoming null ray, is $v\!=\!v_o$, where $v_o$ is determined by the position of the bounce, which in turn can be found inserting $u\!=\!-v$ and $r\!=\!a$ in the last equation. That is
\be
 \left(1-\frac{a}{2m}\right)e^{\frac{a}{2m}}=-v_o^2.
\ee
The bounce time $\tau$ along the $r=R$ worldline is (minus) the Schwarzschild time $t$ of the intersection point multiplied by the red shift factor 
\be
\tau_R=-\sqrt{1-\frac{2m}{R}}\ t.
\ee
The Schwarzschild time in terms of the Kruskal coordinates is given by 
\be
v=\sqrt{\frac{r}{2m}-1}\ e^{\frac{r+t}{4m}}.
\ee
Inserting $r=R$ and $v=v_o$ from the previous equation, we finally get
\be
\tau_R=\sqrt{1-\frac{2m}{R}}\left(R-a-2m\ln{\frac{a-2m}{R-2m}}\right). 
\label{compl1}
\ee
This is a key quantity for our discussion: the bounce time (half the time to the reencounter with the emerging shell), measured by an observer at $R$, given the mass $m$ and the bouncing radius $a$. To study it, let us first take our observer at large radius $R\gg 2m$. Then the above expression simplifies to 
\be
\tau_R \sim (R-a)-2m\ln{\frac{a-2m}{R}}.
\label{approx}
\ee
The term ($R-a)$ is the non relativistic value of the bouncing time. The logarithmic term is the relativistic correction. Something interesting happens when $a\to2m$. The argument of the ln becomes arbitrarily small, therefore \emph{the bouncing time becomes arbitrarily large}:
\be
    \tau_R\underset{a\to 2 m}{\longrightarrow}+\infty.
\ee
Remarkably, this divergence is achievable for any fixed value of the position of the observer $R> 2m$. Hence, as the mirror's extent $a$ approaches the Schwarzschild radius, all observers agree that it takes ``a long time for the process to happen."

Let us discuss the physics of this bouncing time $\tau_R$ in detail, since it is crucial for the following. From the point of view of the observer at the (finite) radius $R$, there is a shell incoming at some time and then a shell coming out an enormous amount of time later.  How so?   

The simple interpretation is in terms of standard time dilation: let us unfreeze the observer's position $R$. Near the bounce, $R\sim a$, the bouncing proper time is of course short; the shell reaches the crystal ball and bounces out always moving at the speed of light.  So the bouncing process is fast, seen locally. But since the bounce happens in a region close to $r=2m$, the slowing down of the local time with respect to an observer far away is huge (as large as $a$ is close to $2m$).  Locally, everything happens fast, but for the observer at $R \gg 2m$ everything happens in slow motion: in terms of his proper time, he sees the shell slowing down (and dimming) while approaching the crystal ball, then lingering a huge amount of time near the mirror, and eventually very slowly the light comes out.  All this, we stress, is standard general relativity.

If $a$ is precisely at $2m$, the waiting time for the light to come out becomes infinite.  What happens is of course that the shell is now so compressed that it generates enough force of gravity to keep the light in.  According to classical general relativity, the light remains trapped forever and a singularity forms.  But this picture cannot be right, because of quantum theory. So, let us step back to $a>2m$ and ask whether and how quantum theory can come into the game.  

To answer, say that $a-2m$ is small and consider an observer at a radius $R$ not much larger than $a$. For this observer we cannot utilise the approximation \eqref{approx} and we must use the complete expression \eqref{compl1} for the bounce time. During the bounce time, the curvature at the observer position is constant in time and is of the order of ${\cal R}\sim m/R^3$.  (For instance the Kretschmann invariant is ${\cal R}^2=R^{abcd} R_{abcd} = \frac{48 m^2}{r^6} \,.$) Since $R>2m$, curvature is small if $m$ is large. We expect local quantum gravity effects to be small in a small curvature region. 

But consider the possibility of a cumulative quantum effect, like in quantum tunnelling or the decay of a radiative atom: the decay probability is small, but if we wait long enough the atom will decay.  Then there is one additional parameter affecting the validity of  the classical theory: the duration of the event. So, the relevant parameter for classicality is, on dimensional grounds,
\be
           q  = l^{2-{b}}_P  \ {\cal R} \ \tau^{b}, 
\ee
with $b$ reasonably taken in the range $b \in [\frac{1}{2}, 2]$. A good guess is $b=1$, for the following reason. A quantum correction of first order in $\hbar$ to the vacuum Einstein equations ${\cal R}icci= 0$  must have the form
\be
   {\cal R}icci + l_p^2  {\cal R}^2 =0. 
\ee
Therefore the force of quantum origin that drives the field away from the classical solution is $\epsilon\,F\sim  l_p^2  {\cal R}^2$. Integrating this in time can give a cumulative effect of the order $l_p^2  {\cal R}^2 t^2$, as for the particle example in the introduction.  Therefore we remain in the classical region only as long as $q\ll 1$, with 
\be
           q  = l_P  \ {\cal R} \ \tau, 
\ee
which corresponds to $b=1$.   Since this heuristic argument is not very strong, we leave ${b}$ undetermined below, to show that our point does not strongly depend on it.

Note that $q$ may become of order unity for $a$ close enough to $2m$ and after a sufficiently long elapsed time. In other words: {\em there is no reason to trust the classical theory outside the horizon for arbitrarily long times and sufficiently close to $r=2m$}.  This is the key conceptual point on which this paper is based.  

Let us see where and when the classical theory can fail. The bounce time $\tau_R$ diverges for any $R$ as $a\to 2m$. The divergence is weak, logarithmic, so for a large mass we need $a$ very close to $2m$ to get $q$ of order unity. Using \eqref{compl1} and the form of the curvature, we have, explicitly
\be
           q  =  \frac{m  l^{2-{b}}_P}{R^3}  \ \left(\sqrt{1-\frac{2m}{R}}\left[R-a-2m\ln{\frac{a-2m}{R-2m}}\right]\right)^{b}. 
\ee
Let us start by inquiring \emph{where} quantum effects are first likely to appear.  This is given by the maximum of $q$ in $R$, in a regime of $a$ near to $2m$.  To find the radius $R_q$ where quantum effects first appear, let us therefore take the derivative of $q$ with respect to $R$ and equate it to zero. After a little algebra, this can be written as:
%\begin{widetext}
\be 
   bR_q^2+[3R_q-(6+b)  m ] \left(a+R_q-2 m
   \ln\frac{a-2 m}{R_q-2 m}\right)=0.
\ee
%\end{widetext}
For small $a-2m$, the logarithmic term dominates, therefore the l.h.s can only vanish if the term in square parenthesis nearly vanishes.  This gives easily the maximum
\be
   R_q=2m \left( 1+\frac{{b}}{6}\right ) +O(a-2m), 
\ee
which is a finite distance, but not much, outside the Schwarzschild radius. This is where quantum effects can first appear.  Notice the nice separation of scales; the result $R_q$ becomes independent of $a$ in the $a \rightarrow 2m$ limit. The quantum effects appear right where they most reasonably should appear: at a macroscopic distance from the Schwarzschild radius, which is necessary for the long bounce time, but close to it, so that the curvature is still reasonably large. 

Let us now compute \emph{when} quantum effects are first likely to appear, at this radius.  Inserting the value of the radius we have found in $q$ we have  
%\begin{widetext}
\be
q=
 \frac{27 (4b)^{\frac{b}{2}}  {l_P}^{2-b}   }{(b +6 )^{3+\frac{b}{2}}m^{2-b}} \left(1+\frac{b}{6}+\frac{a}{2m}- \ln
   \frac{3a-6 m}{b m} \right)^b.
\ee   
%\end{widetext}   

In the limit where $a$ is near $2m$ again it is the ln that dominates and this reduces to 
\be
q=
k\; m^{b-2} {l_P}^{2-b}  \left(-\ln
   \left(a-2 m\right)\right)^b,
\ee   
where $k$ is just a number: $k=27 (4b)^{\frac{b}{2}} / (b +6 )^{3+\frac{b}{2}}$. 
We can have significant quantum effects if $q\sim 1$ namely if 
\be
-\ln \left(a-2 m\right) =l_P^{1-2/b}\frac{m^{2/b-1}
}{k^{1/b}}
\ee 
Inserting this determination of $a$ into the bounce time, we have 
\be
\tau \approx (2{l_p}^{1-\frac2b}k^b)\  m^{\frac2b}. 
\label{compl2}
\ee
In the likely case $b=1$ the quantum effects appear at a distance 
\be
   R=\frac76\ 2m 
\ee
namely a small macroscopic distance outside the Schwarzschild radius, after an asymptotic time
\be
\tau=2k\  \frac{m^2}{l_P}. 
 \label{quantumtime}
\ee

That is: it is possible that quantum gravity affects the \emph{exterior} of the Schwarzschild radius already at a time of order $m^2$. 

Notice that this effect has nothing to do with the $r=0$ singularity: there is no singularity, nor a horizon in the physics considered in this section.  

This is why the argument according to which there cannot be quantum gravity effects outside the horizon, since this region is causally disconnected from the interior of the horizon, is wrong.  In fact, as we have seen, there is room for quantum gravity effects even if there is no interior of the horizon at all.  

We now leave the example, and address the main question of the paper: the construction of the metric of a bouncing hole. 

\section{Time-reversal, Hawking radiation and white holes}\label{III}

General relativity is invariant under the inversion of the direction of time.  This suggests that we can search for the metric of a bouncing star by gluing a collapsing region with its time reversal, where the star is expanding \cite{Hawking2014}.  This is what we shall do.  In doing so, we are going to disregard all dissipative effects, which are not time symmetric.  For instance, the trajectory of a ball that falls down to the ground and then bounces up is time reversion symmetric if we disregard friction, or the inelasticity of the bounce. In a first approximation, disregarding friction and inelasticity, the ball moves up after the bounce precisely in the same manner it fell down.  In the same vein, we disregard all dissipative phenomenon as a first approximation to the bounce of the star.

In particular, we disregard Hawking radiation.  This requires a comment. A widespread assumption is that the energy of a collapsed star is going to be entirely carried away by Hawking radiation.  While the theoretical evidence for Hawking radiation is strong, we do not think that the theoretical evidence for the assumption that the energy of a collapsed star is going to be entirely carried away by Hawking radiation is equally strong. After all, what other physical system do we know where a dissipative phenomenon carries away \textit{all} of the energy of the system? 

Hawking radiation regards the horizon and its exterior: it has no major effect on what happens inside the black hole.  Here we are interested in the fate of the star after it reaches (rapidly) $r=0$. We think that it is also possible to study this physics first, and consider the dissipative Hawking radiation only as  a correction, in the same vein one can study the bounce of a ball on the floor first and then correct for friction and other dissipative phenomena.  This is what we are going to do here. 

Dissipative effects, and in particular the back reaction of the Hawking radiation can then be computed starting  from the metric we construct below. The form given below should be particularly suitable for an analysis of the Hawking radiation using the methods developed by Bianchi and Smerlak \cite{Bianchi2014a,Bianchi2014}, since the map between future and past null infinity needed for this method is entirely coded in the junction functions between spacetime patches.

What should we expect for the metric of the second part of the process, describing the exit of the matter?    The answer is given by our assumption about the time reversal symmetry of the process: since the first part of the process describes the in-fall of the matter to form a black hole, the second part should describe the time reversed process: a white hole streaming out-going matter.

At first this seems surprising.  What does a white hole have to do with the real universe?  But further reflection shows that this is reasonable: if quantum gravity corrects the singularity yielding a region where the classical Einstein equations and the standard energy conditions do not hold, then the process of formation of a black hole does not end in a singularity but continues into the future.  Whatever emerges from such a region is then something that, if continued from the future backwards, would equally end in a past singularity. Therefore it must be a white hole. A white hole solution does not describe something completely unphysical as often declared: instead it is possible that it simply describes the portion of spacetime that emerges from quantum regions, in the same manner in which a black hole solution describes the portion of spacetime that evolves into a quantum region. 

Thus our main hypothesis is that there is a time symmetric process where a star collapses gravitationally and then bounces out.  This is impossible in classical general relativity, because once collapsed a star can never exit its horizon.  Not so if we allow for quantum gravitational corrections.

We make the following assumptions:
\begin{enumerate}[(i)]
\item Spherical symmetry.
\item Spherical shell of null matter: We disregard the thickness of this shell. We use this model for matter because it is simple; we expect our results to generalise to massive matter. In the solution the shell moves in from past null infinity, enters its own Schwarzschild radius, keeps ingoing, enters the quantum region, bounces, and then exits its Schwarzschild radius and moves outwards to infinity.
\item Time reversal symmetry:  We assume the process is invariant under time reversal.
\item Classicality at large radii:  We assume that the metric of the process is a solution of the classical Einstein equations for a portion of spacetime that includes the entire region outside a certain radius, defined below.  In other words, the quantum process is local: it is confined in a finite region of space. 
\item Classicality at early and late times: We assume that the metric of the process is a solution of the classical Einstein equations for a portion of spacetime that includes all of space  before a (proper) time $\epsilon$ preceding the bounce of the shell, and all of space after a (proper) time $\epsilon$ after the bounce of the shell. In other words, the quantum process is local in time: it lasts only for a finite time interval. 
\item No event horizons: We assume the causal structure of spacetime is that of Minkowski spacetime.
\end{enumerate}

\begin{figure}
\includegraphics[height=7.3cm]{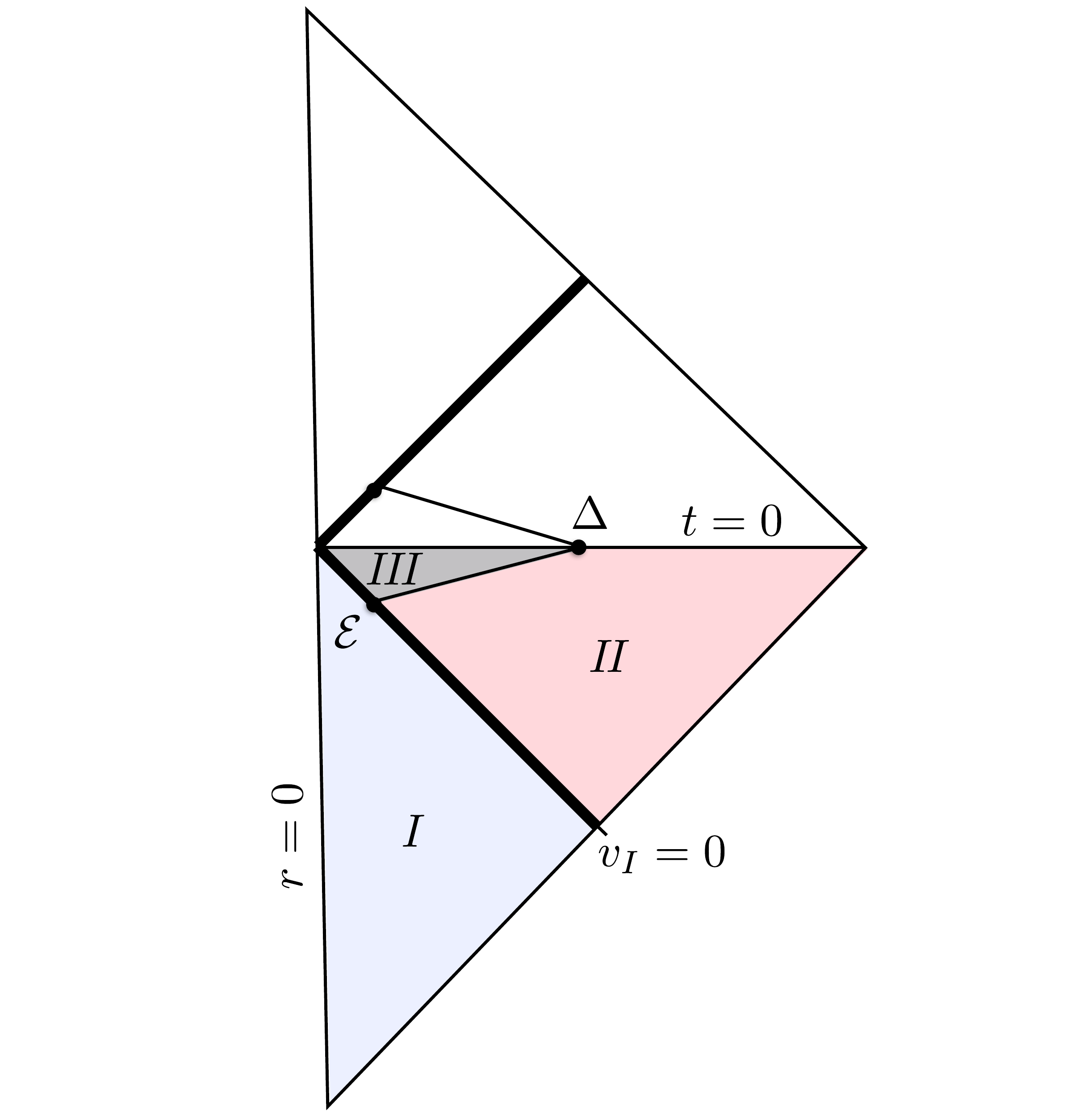}
\vspace{-.5em}
\caption{The spacetime of a bouncing star.}
\label{due}
\end{figure}
This is quite sufficient to our purposes.  

\section{Construction of the bouncing metric}\label{metric}

Because of spherical symmetry, we can use coordinates $(u,v,\theta,\phi)$ with $u$ and $v$ null coordinates in the $r$-$t$ plane and the metric is entirely determined by two functions of $u$ and $v$:
\be
ds^2=-F(u,v) du dv + r^2(u,v)(d\theta^2+\sin^2\theta d\phi^2)
\ee
In the following we will use different coordinate patches, but generally all of this form.  Because of the assumption (vi), the conformal diagram of spacetime is trivial, just the Minkowski one, see Figure \ref{due}.  From assumption (iii) there must be a ``$t=0$" hyperplane which is the surface of reflection of the time reversal symmetry.   It is convenient to represent it in the conformal diagram by an horizontal line as in Figure  \ref{due}. 
Now consider the  incoming and outgoing null shells. By symmetry, the bounce must be at $t=0$. For simplicity we assume (this is not crucial) that it is also at $r=0$. These are represented by the two thick lines at 45 degrees in Figure  \ref{due}.  In the Figure there are two significant points, $\Delta$ and $\cal{E}$, that lie on the boundary of the quantum region. The point $\Delta$ has $t=0$ and is the maximal extension in space of the region where the Einstein equations are violated. Point $\cal{E}$ is the first moment in time where this happens.  We discuss later the geometry of the line joining $\cal{E}$ and $\Delta$. 

Because the metric is invariant under time reversal, it is sufficient for us to construct it for the region below $t=0$ (and make sure it glues well with its future).  The upper region will simply be the time reflection of the lower. The in-falling shell splits spacetime into a region interior to the shell, indicated as $I$ in the Figure and an exterior part.  The latter, in turn, is split into two regions, which we call $II$ and $III$, by the line joining $\cal{E}$ and $\Delta$. Let us examine the metric of these three regions separately: 

\begin{enumerate}[\em (I)]
\item  The first region, inside the shell, must be flat by Bhirkoff's theorem. We denote null Minkowski coordinates in this region $(u_I, v_I, \theta, \phi)$. 
\item The second region, again by Bhirkoff's theorem, must be a portion of the metric of a mass $m$, namely it must be a portion of the (maximal extension of the) Schwarzschild metric. We denote null Kruskal coordinates in this region $(u, v, \theta, \phi)$ and the related radial coordinate $r$. 
\item Finally, the third region is where quantum gravity becomes non-negligible.  We know nothing about the metric of this region, except for the fact that it must join the rest of the spacetime. We denote null coordinates for this quantum region $(u_q, v_q, \theta, \phi)$ and the related radial coordinate $r_q$. 
\end{enumerate}

We can now start building the metric.  Region $I$ is easy: we have the Minkoswki metric in null coordinates determined by 
\be
F(u_I,v_I)=1, \ \ \ \ \ \ \ \ r_I(u_I,v_I)=\frac{v_I-u_I}2.  
\ee
It is bounded by the past light cone of the orgin, that is, by 
\be
v_I= 0.
\ee
In the coordinates of this patch, the ingoing shell is therefore given by $v_I=0$.

\begin{figure}
\includegraphics[height=3.6cm]{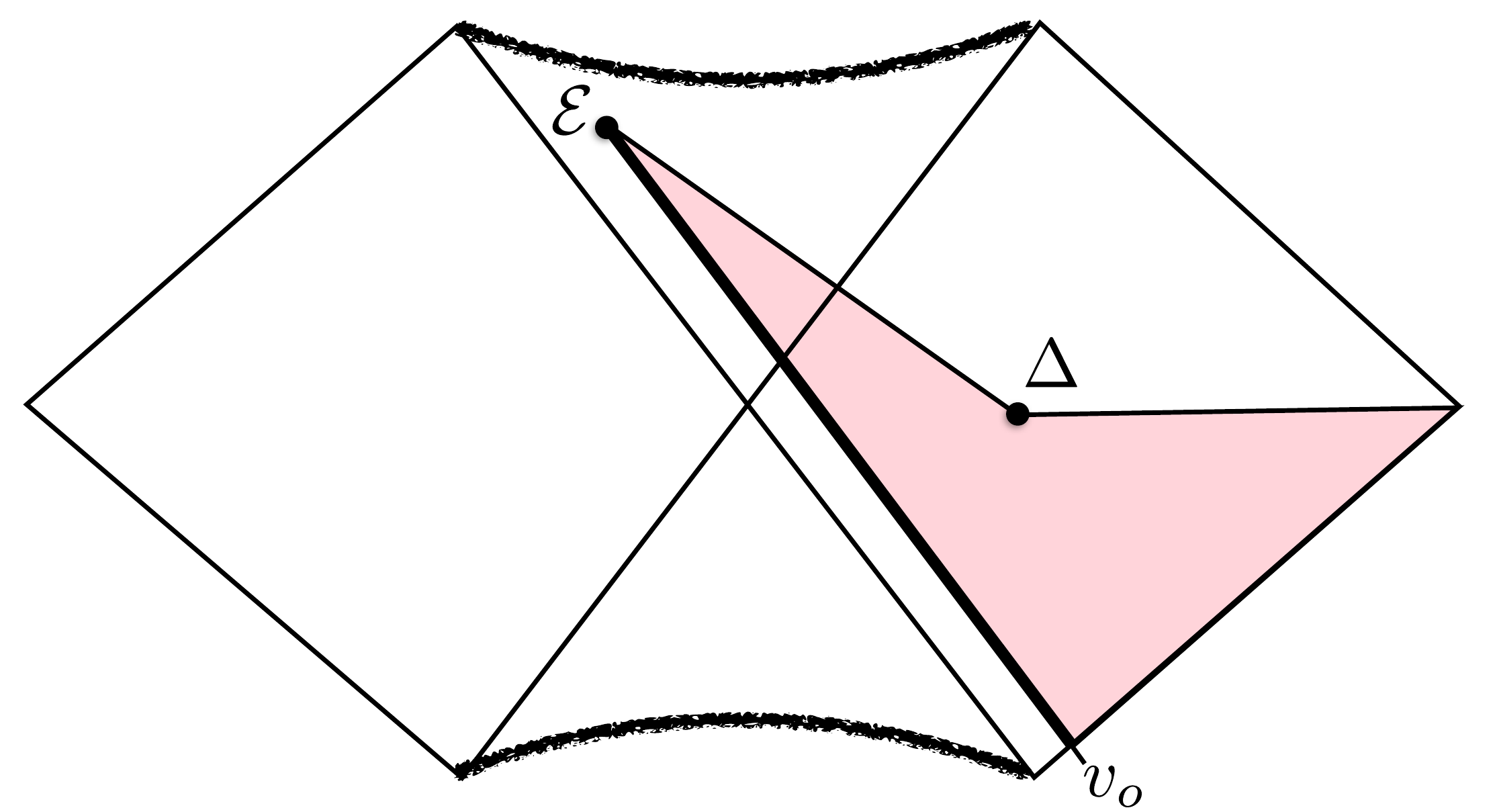}
\caption{Classical black hole spacetime and the region $II$.}
\label{tre}
\end{figure}

Let us now consider region $II$. This must be a portion of the Kruskal spacetime. Which portion?  Put an ingoing null shell in Kruskal spacetime, as in Figure \ref{tre}. The point $\Delta$ is a generic point in the region outside the horizon, which we take on the $t=0$ surface, so that the gluing with the future is immediate. More crucial is the position of the point $\cal{E}$.  Remember that $\cal{E}$ is the point where the in-falling shell reaches the quantum region.  Clearly this must be inside the horizon, because when the shell enters the horizon the physics is still classical. Therefore the region that corresponds to region $II$ in our metric is the shaded region of Kruskal spacetime depicted in Figure \ref{tre}. 

In null Kruskal--Szekeres coordinates the metric of the Kruskal spacetime is given by 
\be
F(u,v)=\frac{32m^3}{r}e^{\frac{r}{2m}}
\ee
with $r$ the function of $(u,v)$ defined by Eq. \eqref{r}.  The region of interest is bounded by a constant $v={v}_o$ null line. The constant $v_o$ cannot vanish, because $v=0$ is an horizon, which is not the case for the in-falling shell.  Therefore $v_o$ is a constant that will enter in our metric.  

The matching between the regions $I$ and $II$ is not difficult, but it is delicate and crucial for the following.  The $v$ coordinates match simply by identifying $v_I=0$ with $v=v_o$. The matching of the $u$ coordinate is determined by the obvious requirement that the radius must be equal across the matching, that is by
\be
r_{I}(u_{I},v_{I}) = r(u,v).
\ee
This gives
\be
\left(1-\frac{v_{I}-u_{I}}{4m}\right)e^{-\frac{v_{I}-u_{I}}{4m}}=u v
\ee
which on the shell becomes
\be
\left(1+\frac{u_{I}}{4m}\right)e^{\frac{u_{I}}{4m}}=u v_o.
\ee
Thus the matching condition is 
\be
u(u_I)=\frac1{v_o}\left(1+\frac{u_{I}}{4m}\right)e^{\frac{u_{I}}{4m}}. 
\ee

\begin{figure}
\includegraphics[height=4.5cm]{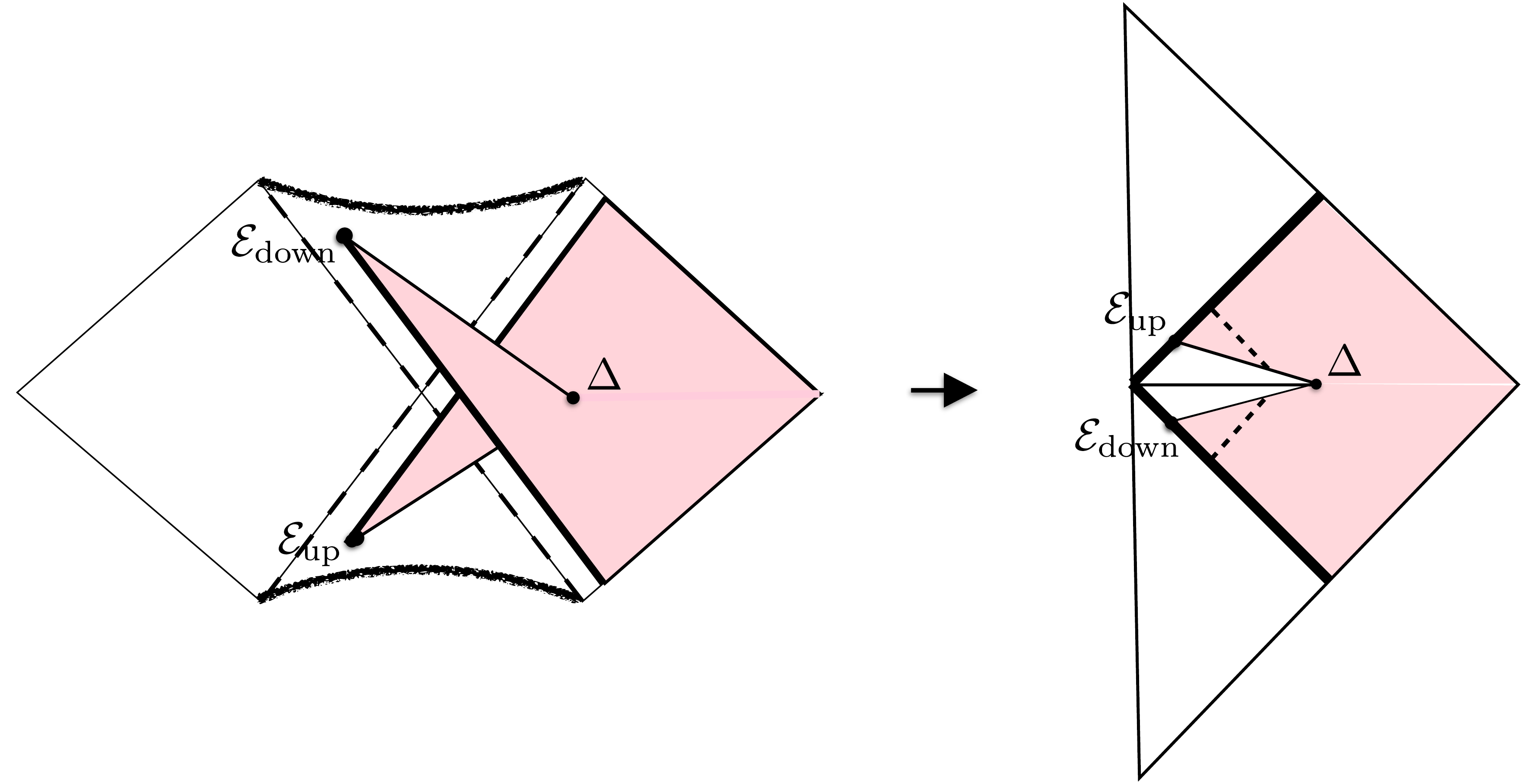}
\caption{The portion of a classical black hole spacetime which is reproduced in the quantum case. The contours $r=2m$ are indicated in both panels by dashed lines.}
\label{four}
\end{figure}

Thus far we have glued the two intrinsic metrics along the boundaries. In order to truly define the metric over the whole region one would also need to specify how tangent vectors are identified along these boundaries, thus ensuring that the extrinsic geometries also matched. However, and perhaps surprisingly, if the induced 3-metrics on the boundaries agree it turns out that it is not necessary to impose further conditions \cite{Barrabes:1991,Dray:1987,Taub:1973}. These works show that the prescription for gluing the tangent spaces is, in that case, uniquely determined.

The matching condition between the region $II$ and its symmetric, time reversed part along the $t=0$ surface is immediate.  Notice, however, that the ensemble of these two regions is not truly a portion of Kruskal space, but rather a portion of a double cover of it, as in Figure \ref{four}: the bouncing metric is obtained by ``opening up" the two overlapping flaps in the Figure and inserting a quantum region in between. 

\begin{figure}
\includegraphics[height=6cm]{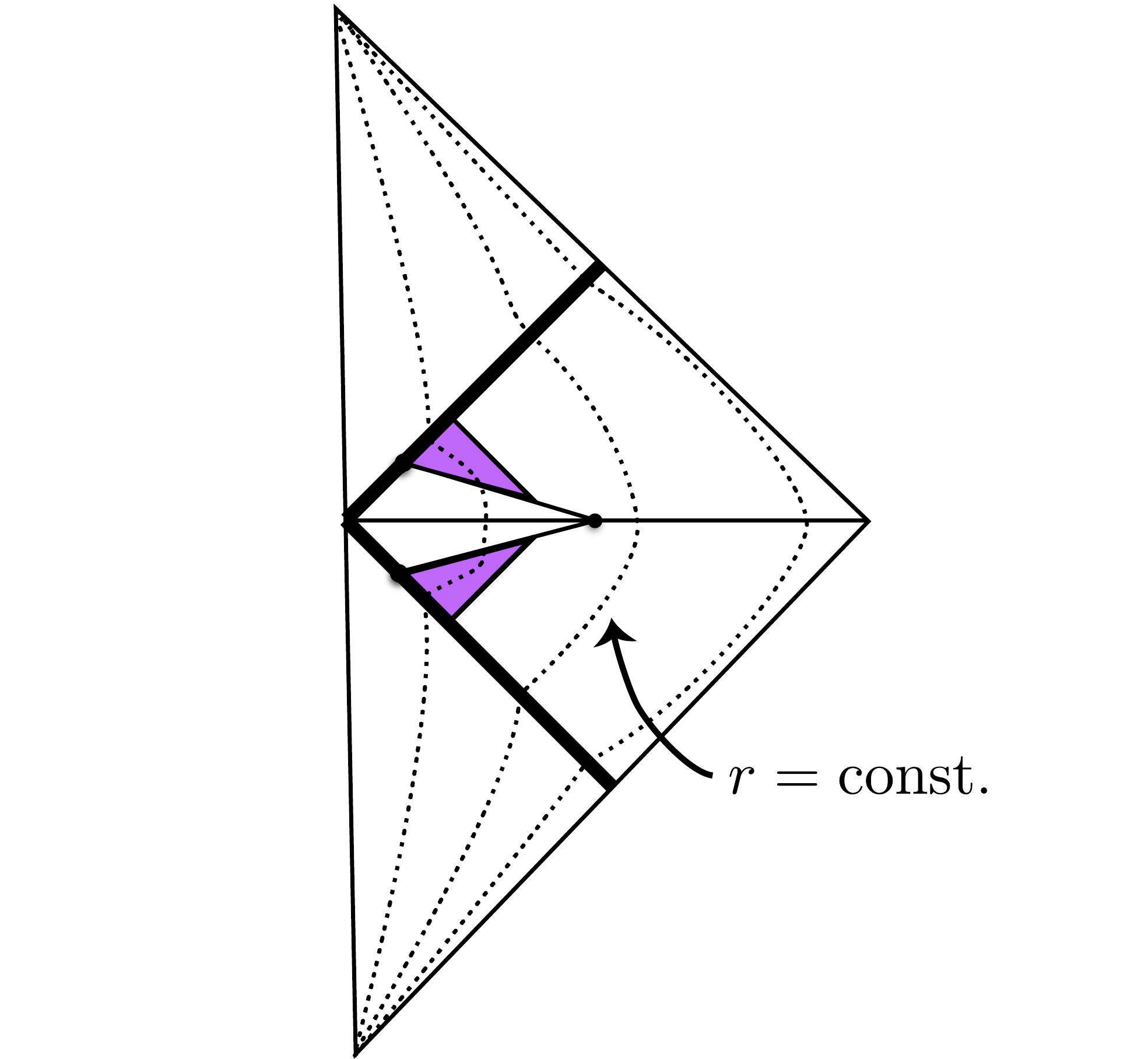}
\caption{Some $r=\text{const.}$ lines. A trapped region is a region where these lines become space-like. There are two trapped regions in this metric, indicated by shading.}
\label{five}
\end{figure}

It remains to fix the points $\cal{E}$ and $\Delta$, the line connecting them and the metric of the quantum region.   We take $\cal{E}$ to be the point that has $(u_I,v_I)$ coordinates $(-2\epsilon, 0)$ and $\Delta$ the point that has Schwarzschild radius $r=2m+ \delta $ and lies on the time reversal symmetry line $u+v=0$. Here $\epsilon$ and $\delta$ are two constant with dimensions of length that determine the metric. Lacking a better understanding of the quantum region, we take the line connecting $\cal{E}$ and $\Delta$ to be the (spacelike) geodesic between the two.
Finally, we fix the metric in the region $III$ as follows. We use $(u_{q},v_{q})$ that are equal to the  $(u,v)$ coordinates on the boundary and choose simply 
\be
F(u_{q},v_q)=\frac{32m^3}{r_q}e^{\frac{r_q}{2m}}, 
\ee
where $r_{q}$ is the function of $(u_{q},v_{q})$ 
\be
r_{q}=\frac{1}{2}(v_q-u_{q}). 
\ee
This is only a simple first ansatz, to be ameliorated as understanding of this region and of quantum gravity improves.  What is important is that the $r_q=\text{const.}$ surfaces are again timelike in region $III$. Therefore region $III$ is outside the trapped region.  The trapped region is bounded by the incoming shell trajectory, the null $r=2m$ horizon in the region $II$, and the boundary between region $II$ and region $III$. The two trapped regions are depicted in Figure \ref{five}. \\

This concludes the construction of the metric, which is now completely defined. It satisfies all the requirements with which we began. It describes, in a first approximation and disregarding dissipative effects, the full process of gravitational collapse, quantum bounce and explosion of a star of mass $m$.  It depends on four constants: $m,v_o, \delta, \epsilon$, whose physical meaning will be discussed below. In the following sections we study some of its properties.

\section{Exterior time, interior time}\label{time}

Consider two observers, one at the center of the system, namely at $r=0$, and one that remains at radius $r=\r > 2m$. In the distant past, both observers are in the same Minkowski space.  Notice that the entire process chooses a Lorentz frame: the one where the center of mass of the shell is not moving. Therefore the two observers can synchronise their clocks in this frame.   In the distant future the two observers find themselves again in a common Minkowski space with a preferred frame and therefore can synchronise their clocks again. However, there is no reason for the proper time $\tau_o$ measured by one observer to be equal to the proper time $\tau_R$ measured by the other one, because of the conventional, general relativistic time dilation. Let us compute the time difference accumulated between the two clocks during the full process. 

The two observers are both in a common Minkowski region until the shell reaches $\r$ while falling in and they are again both in this region after the shell reaches $\r $ while going out. In the coordinate system $(t_I=(v_I+u_I)/2,\ r_I=(v_I-u_I)/2)$ of the region $I$, these are the points with coordinates $(-\r ,\r )$ and $(\r ,\r )$ respectively. The two simultaneous points for the inertial observer at $r=0$ are 
$(-\r,0)$ and $(\r,0)$ and his proper time is clearly 
\be
	\tau_{0}=2\r.
\ee

Meanwhile, the observer at $r=\r$ sits at constant radius in a Schwarzschild geometry.  The proper time between the two moments she crosses the shell is twice the time from the first crossing to the $t=0$ surface. This is analogous to twice the bounce time we have computed in Section \ref{ball}, but let us redo the calculation here, to avoid confusion, since the overall context is different (the relevant parameter is $v_o$ rather than $a$). Since the observer is stationary, the proper time is given by
\be
\tau_{\r } = -2 \left(1-\frac{2m}{\r }\right)^{\frac{1}{2}} t,
\ee
where $t$ is the Schwarzschild time. Therefore the proper time can simply be found by transforming the coordinates $(u,v)$ to Schwarzschild coordinates. The standard change of variables to the Schwarzschild coordinates in the exterior region $r>2m$ is
\be
\frac{u+v}2=\left(\frac{r}{2m}-1\right)^{\frac12} e^{\frac{r}{4m}}\sinh\frac{t}{4m},
\ee
\be
\frac{v-u}2=\left(\frac{r}{2m}-1\right)^{\frac12} e^{\frac{r}{4m}}\cosh\frac{t}{4m}.
\ee
Along the shell's in-fall $v=v_o$ and so 
\be
t = 4m \ln \left( \frac{v_o}{\left( \frac{\r }{2m}-1 \right)^{1/2}} e^{- \frac{\r  }{4m}} \right).
\ee
Therefore the total time measured by the observer at radius $\r$ is 
\be
	\tau_{\r}= \sqrt{1- \frac{2m}{\r} } \left( 2\r  -8m \ln v_o + 4m \ln \frac{\r-2m }{2m}  \right).
\ee
If the external observer is at large distance, $R\gg 2m$, we obtain, to the first relevant order, the difference in the duration of the bounce measured outside and measured inside to be
\be
	\tau = \tau_{\r}- \tau_o  =- 8m \ln v_o.
	\label{vo}
\ee
This can be arbitrarily large as $v_o$ is arbitrarily small.   

The process seen by an outside observer takes a time arbitrarily longer than the process measured by an observer inside the collapsing shell.   

In the next Section we determine $v_o$, and therefore the duration of the bounce seen from the outside. 

\section{The constants of the metric and the breaking of the semiclassical approximation}\label{constants}

The metric we have constructed depends on the mass $m$ and three additional constants: $v_o, \epsilon, \delta$. We now determine all of them as functions of $m$.  

The constant $\epsilon$ fixes the moment in which the collapsing shell abandons the region where the classical theory is reliable.  In quantum gravity, we exit the quantum region when the matter density, or the curvature, reaches the Planck scale (see a full discussion in \cite{Ashtekar:2006es}).  This must also be true for black holes \cite{Frolov:1979tu,Hayward2006,frolov:BHclosed,Bardeen2014,Rovelli2014}. The curvature $\cal R$ is of the order $m/r^3$ and reaches the Planck value ${\cal R}\sim l_P^{-2}$ when 
\be
r\sim (m l_P^{2})^{\frac13}=\left(\frac{m}{m_P^3}\right)^{\frac13} l_P.
\ee
Here $l_P$ and $m_P$ are the Planck length and the Planck mass.  Therefore we expect the parameter $\epsilon$ to be of the order
\be
\epsilon \sim \left(\frac{m}{m_P^{3}}\right)^{\frac{1}{3}} l_P.
\ee

The parameter $\delta$ is the most important of all. To understand its meaning, consider the quantum region $III$. A part of it is inside $r=2m$. This is very reasonable, since this part surrounds the region where the classical singularity would appear.  However, a part of the region $III$ leaks \emph{outside} the $r=2m$ sphere.  This is needed, if we want to avoid the event horizon and have the bounce, because if the entire region $r\ge 2m$ were classical, an event horizon would be unavoidable, as the $r=2m$ classical surface is null. If an event horizon forms matter cannot bounce out and a singularity is unavoidable. 

Thus the quantum effect must leak outside $r=2m$.  We have shown in Section \ref{ball} that this can happen without violating the validity of the semiclassical approximation, because of the piling up of corrections. But we have also seen that for this to happen we need a long time, which we have estimated in Section \ref{ball} to be given by equation  \eqref{quantumtime}. In order for the process to last this long, $v_o$ must be small. Indeed, we have seen in the previous section that the duration of the process is determined by $v_o$ via equation \eqref{vo}. Bringing the two together we find the condition
\be
	\tau=-8m \ln v_o> \tau_{q} = 4k\, \frac{m^2}{l_p},
\ee
that is
\be
	v_o<e^{-k\, \frac{m}{2l_p}},
\ee
which is very small for a macroscopic black hole.  Let us therefore fix $v_o$ to the value $v_o=e^{-k\, \frac{m}{2l_p}}$ that minimises the bounce time and yet still yields a sufficiently long  time for quantum gravity to act. In turn, this fixes $\delta$, because $\delta$ is bounded from below by $v_o$. The value of $\delta$ can easily be deduced from the discussion in Section \ref{ball}: the quantum region needs to extends all the way to $7/6$'th of the Schwarzschild radius. That is, $2m+ \delta = \frac{7}{6}(2m)$ or
\be
    \delta=\frac{m}{3}.
\ee
Notice that $\delta $ is of the order of the size of the black hole itself. 

Summarising, the metric we have constructed  is determined by a single constant: the mass $m$ of the collapsing shell.  The other constants are fixed in terms of the mass and the Planck constants. 
\begin{eqnarray}
\epsilon &\sim& \left(\frac{m}{m_P^{3}}\right)^{\frac{1}{3}} l_P,\\
v_o&\sim&e^{-k\, \frac{m}{2l_p}},\\
    \delta&\sim&\frac{m}{3}. 
\end{eqnarray}
A tentative time reversal symmetric metric describing the quantum bounce of a star is entirely defined.   \\

\section{Relation with a full quantum gravity theory}\label{conclusion}

We have constructed the metric of a black hole tunnelling into a white hole by using the classical equations outside the quantum region, an order of magnitude estimate for the onset of quantum gravitational phenomena, and some indirect indications on the effects of quantum gravity. This, of course, is not a first principle derivation.  For a first principle derivation a full theory of quantum gravity is needed.

However, the metric we have presented poses the problem neatly for a quantum gravity calculation.  The problem now can be restricted to the calculation of a quantum transition in a finite portion of spacetime.   

The quantum region that we have determined is bounded by a well defined classical geometry. Given the classical boundary geometry, can we compute the corresponding quantum transition amplitude?  Since there is no classical solution that matches the in and out geometries of this region, the calculation is conceptually a rather standard tunnelling calculation in quantum mechanics. 

Indeed, this is precisely the form of the problem that is adapted for a calculation in a theory like covariant loop quantum gravity \cite{Rovelli2011c,Rovelli}. The spinfoam formalism is designed for this. Notice that the process to be considered is a process that takes a short time and is bounded in space. Essentially, we want to know the transition probability between the state with the metric on the lower to upper $\cal{E}$-$\Delta$ surfaces. This may be attacked for instance, in a vertex expansion, to first order. If this calculation can be done, we should then be able to replace the order of magnitudes estimates used here with a genuine quantum gravity calculation. And, in particular, compute from first principles the duration $\tau$ of the bounce seen from the exterior. We leave this for the future. 

\acknowledgments

CR thanks Steve Giddings for a fruitful exchange. He also thanks Don Marolf and Sabine Hossenfelder for very useful discussions. Both authors thank Xiaoning Wu for discussions on gluing metrics. HMH acknowledges support from the National Science Foundation (NSF) International Research Fellowship Program (IRFP) under Grant No. OISE-1159218.

\bibliographystyle{utcaps}
\bibliography{library}

\providecommand{\href}[2]{#2}\begingroup\raggedright\begin{thebibliography}{10}

\bibitem{Narayan2013}
R.~Narayan and J.~E. McClintock, ``{Observational Evidence for Black Holes},''
  \href{http://arxiv.org/abs/1312.6698}{{\tt arXiv:1312.6698}}.

\bibitem{Smolin1997}
L.~Smolin, {\em {The Life of the Cosmos}}.
\newblock Oxford University Press, 1997.

\bibitem{Frolov:1979tu}
V.~P. Frolov and G.~A.~Vilkovisky, ``{Quantum Gravity removes Classical
  Singularities and Shortens the Life of Black Holes},''
ICTP preprint IC/79/69, Trieste (1979).

\bibitem{Frolov:1981}
V.~P. Frolov and G.~A.~Vilkovisky, ``Spherically symmetric collapse in quantum gravity,"
{\em Phys. Lett. B} {\bf 106} (1981) 307.

\bibitem{Stephens1994}
C.~R. Stephens, G.~t. Hooft, and B.~F. Whiting, ``{Black hole evaporation
  without information loss},''{\em Classical and Quantum Gravity} {\bf 11}
  (Mar., 1994)  621--647, \href{http://arxiv.org/abs/9310006}{{\tt
  arXiv:9310006 [gr-qc]}}.

\bibitem{Mazur:2004}
P.~O.~Mazur and E. ~Mottola, 
  ``Gravitational vacuum condensate stars,"
  {\em Proc. Nat. Acad. Sci. U.S.A.} {\bf 101} (2004) 9545Ð50.

\bibitem{Ashtekar:2005cj}
A.~Ashtekar and M.~Bojowald, ``{Black hole evaporation: A paradigm},'' {\em
  Class. Quant. Grav.} {\bf 22} (2005)  3349--3362,
  \href{http://arxiv.org/abs/0504029}{{\tt arXiv:0504029 [gr-qc]}}.

\bibitem{Hayward2006}
S.~A. Hayward, ``{Formation and Evaporation of Nonsingular Black Holes},'' {\em
  Phys. Rev. Lett.} {\bf 96} (2006)  031103,
  \href{http://arxiv.org/abs/0506126}{{\tt arXiv:0506126 [gr-qc]}}.

\bibitem{Hossenfelder:2010}
S.~Hossenfelder, L.~Modesto, and I.~Premont-Schwarz,  ``A model for non-singular black hole collapse and evaporation"
{\em Phys. Rev. D} {\bf 81}  (2010) 44036.
\href{http://arxiv.org/abs/0912.1823}{\tt arXiv:0912.1823}

\bibitem{frolov:BHclosed}
V.~P. Frolov, ``{Information loss problem and a ``black hole'' model with a
  closed apparent horizon},'' \href{http://arxiv.org/abs/1402.5446}{{\tt
  arXiv:1402.5446}}.

\bibitem{Bardeen2014}
J.~M. Bardeen, ``{Black hole evaporation without an event horizon},''
  \href{http://arxiv.org/abs/1406.4098}{{\tt arXiv:1406.4098}}.

\bibitem{Rovelli2014}
C.~Rovelli and F.~Vidotto, ``{Planck stars},''
  \href{http://arxiv.org/abs/1401.6562}{{\tt arXiv:1401.6562}}.

\bibitem{Barrau2014}
A.~Barrau and C.~Rovelli, ``{Planck star phenomenology},''
  \href{http://arxiv.org/abs/1404.5821}{{\tt arXiv:1404.5821}}.

\bibitem{Giddings:1992}
S.~Giddings, ``Black holes and massive remnants,"
{\em Physical Review D} {\bf 46} (1992) 1347, 
\href{http://arxiv.org/abs/hep-th/9203059}{\tt arXive:hep-th/9203059}.

\bibitem{HAJICEK2001}
P.~H\'{a}j\'{\i}\v{c}ek and C.~Kiefer, ``{Singularity avoidance by collapsing
  shells in quantum gravity},''{\em International Journal of Modern Physics D}
  {\bf 10} (Dec., 2001)  775--779, \href{http://arxiv.org/abs/0107102}{{\tt
  arXiv:0107102 [gr-qc]}}.

\bibitem{Ashtekar2006}
A.~Ashtekar, T.~Pawlowski, and P.~Singh, ``{Quantum Nature of the Big
  Bang},''{\em Physical Review Letters} {\bf 96} (Apr., 2006)  141301,
  \href{http://arxiv.org/abs/0602086}{{\tt arXiv:0602086 [gr-qc]}}.

\bibitem{Ashtekar:2006es}
A.~Ashtekar, T.~Pawlowski, P.~Singh, and K.~Vandersloot, ``{Loop quantum
  cosmology of k=1 FRW models},'' {\em Phys. Rev.} {\bf D75} (2007)  24035,
  \href{http://arxiv.org/abs/0612104}{{\tt arXiv:0612104 [gr-qc]}}.

\bibitem{Rovelli2013e}
C.~Rovelli and E.~Wilson-Ewing, ``{Why are the effective equations of loop
  quantum cosmology so accurate?},'' \href{http://arxiv.org/abs/1310.8654}{{\tt
  arXiv:1310.8654}}.

\bibitem{Almheiri:2012rt}
A.~Almheiri, D.~Marolf, J.~Polchinski, and J.~Sully, ``{Black Holes:
  Complementarity or Firewalls?},'' {\em JHEP} {\bf 1302} (2013)  62,
  \href{http://arxiv.org/abs/1207.3123}{{\tt arXiv:1207.3123}}.

\bibitem{Hawking2014}
S.~W.~Hawking, ``Information Preservation and Weather Forecasting for Black Holes," 
\href{}{\tt arXiv:1401.5761}

\bibitem{Bianchi2014a}
E.~Bianchi and M.~Smerlak, ``{Entanglement entropy and negative-energy fluxes
  in two-dimensional spacetimes},'' \href{http://arxiv.org/abs/1404.0602}{{\tt
  arXiv:1404.0602}}. \url{http://arxiv.org/abs/arXiv:1404.0602}.

\bibitem{Bianchi2014}
E.~Bianchi and M.~Smerlak, ``{Last gasp of a black hole: unitary evaporation
  implies non-monotonic mass loss},''
  \href{http://arxiv.org/abs/1405.5235}{{\tt arXiv:1405.5235}}.

\bibitem{Barrabes:1991}
C. ~Barrab\`es and W.~Israel, 
  ``{Thin shells in general relativity and cosmology: The lightlike limit},"
 {\em Phys. Rev. D} {\bf 43} (1991) 1129.

\bibitem{Dray:1987}
C.~J.~S. Clarke and T.~Dray,
``Junction conditions for null hypersurfaces," 
{\em Class. Quantum Grav.} {\bf 4} (1987) 265.

\bibitem{Taub:1973}
A.~H.~Taub,
{\em Commun. Math. Phys. } {\bf 29} (1973) 79.

\bibitem{Rovelli2011c}
C.~Rovelli, ``{Zakopane lectures on loop gravity},'' {\em PoS} {\bf QGQGS2011}
  (2011)  3, \href{http://arxiv.org/abs/1102.3660}{{\tt arXiv:1102.3660}}.

\bibitem{Rovelli}
C.~Rovelli and F.~Vidotto, {\em {Introduction to covariant loop quantum
  gravity}}.
\newblock Cambridge University Press, to appear., 2015.

\end{thebibliography}\endgroup

\end{document}